\documentclass[twocolumn,showpacs,floatfix,prl]{revtex4}
\usepackage{graphicx}
\usepackage{dcolumn}
\usepackage{bm}
\usepackage{amsmath}
\usepackage{amssymb}

\begin{document}

\noindent {\bf Comment on "Spectral Signatures of the Fulde-Ferrell-Larkin-Ovchinnikov Order Parameter in
One-Dimensional Optical lattices"}

In a recent letter Reza Bakhtiari {\it et al.} \cite{Torma}
studied an imbalanced two-component atomic Fermi gas in a
one-dimensional (1D) optical lattice with a trapping
potential, within the Bogoliubov-de Gennes (BdG) approximation.
They showed that the prominent oscillations of the pairing gap
(within the BdG approximation), characteristic of a
Fulde-Ferrell-Larkin-Ovchinnikov state (FFLO), could be detected
in the rf spectra and in the momentum-resolved photoemission
spectra of the gas.  In this comment we show that the BdG
approximation not only produces inaccurate results for the
examples presented in \cite{Torma}, but that they are
qualitatively incorrect making the analysis of the rf spectra
unreliable and shedding doubts on the applicability of rf
spectroscopy to detect the FFLO state in 1D optical lattices.

The Letter \cite{Torma} is devoted to 1D optical
lattices without any reference to higher dimensions where the BdG
approximation could be valid. The use of the BdG approximation is
justified by saying that it provides qualitative information on
the system and allows one to calculate the rf spectrum. In spite of
the known failure of the BdG approximation in 1D systems, it
 is still used \cite{Torma2}.
In the absence of a trap the BCS treatment gives large errors in
the order parameter as compared to the exact solution
(see Ref. 19 of the letter). With the inclusion of the trap we
have to resort to the Density Matrix Renormalization Group (DMRG)
to produce numerically "exact" solutions for 1D lattice problems
of moderate sizes ($\sim 100$ sites). Indeed, the DMRG has already been
used to study the FFLO phase in a 1D trapped lattice gas (see Ref.
11 and 16 of the letter). Therefore the DMRG constitutes an ideal
benchmark to test the accuracy of the BdG approximation in 1D
lattice problems.

\begin{figure}[ht]
\includegraphics[angle=0,width=9.0cm,scale=1.0]{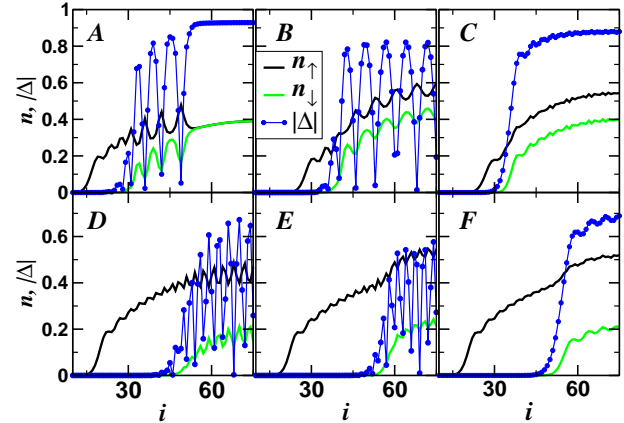}
\caption{(Color online) Density of majority and minority atoms $n_i$ and absolute value of the local gap parameter
$|\Delta_i|$ as a function of the site index $i$. Upper panels show results for $P=0.23$ and lower panels for
$P=0.70$. From left to right we show results of BdG without the Hartree term, BdG with the Hartree term and DMRG.
 } \label{fig1}
\end{figure}

In Fig. 1 we show the densities and the absolute value of the gap for a system with
 polarization $P=0.23$ ($N_{\uparrow}=40$, $N_{\downarrow}=25$) in the upper panels and for $P=0.70$
 ($N_{\uparrow}=40$, $N_{\downarrow}=7$) in the lower panels, in
a lattice of $L=150$ sites. Panels $A$ and $D$ correspond to the
BdG approximation neglecting the Hartree term like in Fig. 1 of
\cite{Torma}. In panels $B$ and $E$ include the Hartree term into
the BdG approximation. Panels $C$ and $F$ are DMRG results. The
gap parameter is defined as $|\Delta_i|=U \sqrt{ ( <n_{i \uparrow}
n_{i \downarrow}> - <n_{i \uparrow}> <n_{i \downarrow}>)}$, where
the expectation values are taken in the ground state wavefunctions
of the three approximations. Within the BdG approximations this
definition coincides with local pairing gap. The Hartree term,
neglected in \cite{Torma}, does modify the results
towards the "exact" DMRG results. However, both BdG approximations
are still quite far from the DMRG results, specially for lower
polarizations. The most dramatic differences can be seen in the
pairing gap with very large amplitude oscillations which are
completely softened in the DMRG results. It might be argued that the
dimensions of the systems considered are small, however, doubling
the size of the systems the authors found similar results as we
did using DMRG.

The subsequent analysis of the rf spectra is based on the BdG results of panels $A$ and $D$. Taking into account
that the gap oscillation are reduced by one order of magnitude in the DMRG results, it is doubtful that rf
spectroscopy could provide information about the spatial structure of the pairing gap. 
Whether rf spectroscopy could signal the FFLO phase in 1D lattice
systems is still an open question which might be confirmed by rf experiments,
or addressed numerically by means of a DMRG study. Note that within the Correction Vector Approach
the calculation of the excitation spectrum is not needed.

This work is supported by Spanish grants  FIS2006-12783-C03-01 and
CAM-CSIC CCG07-CSIC/ESP-1962.

\medskip

\noindent R. A. Molina$^1$, J. Dukelsky$^1$, and  P. Schmitteckert$^2$
\indent $^1$Instituto de Estructura de la Materia, CSIC,
\indent Serrano 123, 28006 Madrid, Spain \\
\indent $^2$Institut f\"ur Nanotechnologie, Forschungzentrum
\indent Karlsruhe, 76021 Karlsruhe, Germany

\medskip

\noindent {\bf PACS numbers:} 03.75.Ss, 74.45.+c, 78.90.+t

\end{document}